\documentclass[preprint,12pt]{elsarticle}

\usepackage{graphicx}
\usepackage{mathrsfs}
\usepackage{amssymb}
\usepackage{amsmath}
\usepackage{float}
\usepackage{color}
\usepackage{subfigure}
\usepackage{mathtools}
\usepackage[colorlinks=true,linkcolor=blue,urlcolor=black,citecolor=blue]{hyperref}
\newcommand{\red}[1]{{\textcolor{red}{#1}}}
\newcommand{\bra}[1]{\langle #1 \vert}

\newcommand{\ket}[1]{\vert #1 \rangle}
\newcommand{\be}{\begin{equation}}
\newcommand{\ee}{\end{equation}}
\newcommand{\beq}{\begin{eqnarray}}
\newcommand{\eeq}{\end{eqnarray}}
\newcommand{\la}{\langle}
\newcommand{\ra}{\rangle}
\newcommand{\nn}{\nonumber}
\newcommand{\commentold}[1]{}
\begin{document}

\begin{frontmatter}


\title{The Mpemba effect in quantum oscillating and two-level systems}

\ead{f.kheirandish@uok.ac.ir}
\author[a]{F. Kheirandish}
\author[a]{N. Cheraghpour}
\author[b]{A. Moradian}
\affiliation[a]{organization={ Department of Physics, University of Kurdistan},
            addressline={Pasdaran Boulevard},
            city={Sanandaj},
            postcode={P.O.Box 66177-15175},
            state={Kurdistan},
            country={Iran}}
\affiliation[b]{organization={Department of Physics, University of Garmian},
            addressline={Garmian},
            city={Bardesur},
            postcode={M923+GRW},
            state={Kalar},
            country={Iraq}}

\begin{abstract}
The Mpemba effect (ME) is investigated in the context of ubiquitous quantum oscillators and two-level systems (TLS) using a novel approach. Exact reduced density matrices for various initial states are derived. The temporal behavior of the trace distance for these initial states is calculated analytically and presented. For a dissipative quantum oscillating system, it is demonstrated that number states $|N\ra$ intersect with coherent states $|\alpha\ra$, with this intersection occurring earlier for smaller values of $N$. Additionally, thermal states intersect with coherent states for specific values of $|\alpha|$, leading to the occurrence of the ME in these two scenarios. A weaker version of the ME is also observed for thermal and number states. In the case of a quantum TLS, it is shown that the ME occurs, and the potential for its experimental observation is discussed, with reference to the Jaynes-Cummings model (JCM) involving a time-dependent coupling function.
\end{abstract}

\end{frontmatter}

\section{Introduction}
The Mpemba effect (ME) \cite{Mpemba1969,Kell1969,Jeng2006} refers to the counterintuitive phenomenon wherein, under specific conditions, a system initially farther from equilibrium can approach equilibrium more rapidly than one that is initially closer to it. A well-known and longstanding illustration of this phenomenon is the surprising observation that, in some cases, warmer water can freeze more quickly than cooler water. The ME has been the subject of considerable scientific debate for decades \cite{Burridge2016,Bechhoefer2021} and, despite extensive investigations in various classical systems \cite{Jeng2006,Burridge2016,Bechhoefer2021,Lasanta2017,Lapolla2020,Lapolla2019,Kumar2020,Biswas2023}, it remains an intriguing and partially unresolved phenomenon that continues to capture the interest of researchers. In recent years, significant attention has been directed toward studying the ME at the microscopic scale \cite{Nava2019,Manikandan2021,Carollo2021,Chatterjee2023,Ares2023,Ivander2023, Joshi2024,Rylands2024,Chatterjee2024,Zhang2024,Yamashika2024,Shapira2024,Liu2024,Warring2024}, with both theoretical predictions and experimental confirmations of its quantum manifestations. These developments are driving efforts to uncover Mpemba-like behaviors rooted in quantum mechanics, which may hold significance for the field of quantum thermodynamics and offer potential applications in advanced technologies, including quantum computing and energy storage \cite{Warring2024}. The ME has been observed across a broad array of physical systems, such as clathrate hydrates \cite{Ahn2024}, polymers \cite{Hu2024}, magnetic alloys \cite{Chaddah}, carbon-nanotube resonators \cite{Greaney2011}, and dilute atomic gases \cite{Keller2018}. A major advancement in the theoretical understanding of this effect emerged from analyses grounded in the stochastic thermodynamics of Markovian processes \cite{Raz2017,Klich2019}. Recent studies predict the emergence of the ME in the realm of photonics \cite{Longhi2024_1,Longhi2024_2}. In \cite{Nava2024} models coupled to multiple reservoirs and the possibility of observing the ME using electron currents have been investigated. A strong ME has been reported in \cite{Xia2025} by engineering the rotation in the initial state and applying efficient gate operations on a single trapped ion in its ground state. The ME in  general non-Markovian open quantum systems has been investigated in \cite{Strachan2025}. For comprehensive reviews on the ME, the interested reader is referred to \cite{Ares2025, Teza2025}.

Here, we explore ME in the context of ubiquitous quantum oscillator and two-level systems (TLS) using a new methodology \cite{Kh2024}. Exact reduced density matrices for various initial states are derived, and the temporal behavior of the trace distance for these states is calculated analytically. In a dissipative quantum oscillator, it is shown that number states $|N\ra$ intersect with coherent states $|\alpha\ra$ with the intersection occurring sooner for smaller values of $N$. Additionally, thermal states intersect with coherent states at specific values of $|\alpha|$, resulting in the ME occurring in these two cases. A milder form of the ME is also observed for thermal and number states. For a quantum TLS, the ME is found to occur, and the potential for its realization and experimental detection is discussed, with reference to the Jaynes-Cummings model (JCM) \cite{Gerry2012} involving a time-dependent coupling function.
\section{Dissipative Harmonic Oscillator}\label{sec2}
\subsection{A novel scheme}
As our initial example, we consider the ubiquitous quantum harmonic oscillator interacting with a bosonic heat bath characterized by the inverse temperature $\beta_b$. The bath oscillator mirrors the main oscillator and is described by the ladder operators $\hat{b},\,\hat{b}^\dag$. These two oscillators interact through a time-dependent coupling function $g(t)$. The Hamiltonian is given by
\begin{equation}\label{H1}
\hat{H}(t)=\hbar\omega_0\,\hat{a}^\dag\hat{a}+\hbar\omega_0\hat{b}^\dag\hat{b}+\hbar g(t)\,(\hat{a}\hat{b}^\dag+\hat{a}^\dag\hat{b}).
\end{equation}
The number operators corresponding to $a$ and $b$-oscillators are defined as $\hat{n}_a=\hat{a}^\dag\hat{a}$ and $\hat{n}_b=\hat{b}^\dag\hat{b}$, respectively. Using the Bogoliubov transformations
\begin{eqnarray}\label{Bogo}
\hat{a} &=& \frac{1}{\sqrt{2}}(\hat{A}+\hat{B}),\nonumber\\
\hat{b} &=& \frac{1}{\sqrt{2}}(\hat{B}-\hat{A}),
\end{eqnarray}
the Hamiltonian becomes separable in terms of the new ladder operators
\begin{equation}\label{HAB}
\hat{H}(t)=\hbar\omega_A (t)\,\hat{A}^\dag\hat{A}+\hbar\omega_B (t)\,\hat{B}^\dag\hat{B},
\end{equation}
where we have defined $\omega_A (t)=\omega_0-g(t)$, $\omega_B (t)=\omega_0+g(t)$. For later convenience, we also define $\kappa(t)=\int\limits_0^t g(t')dt'$. The time-evolution operator corresponding to Hamiltonian Eq.~(\ref{H1}) is
\begin{equation}\label{U1}
\hat{U} (t)=e^{-i\Omega_A (t)\,\hat{A}^\dag\hat{A}}\otimes e^{-i\Omega_B (t)\,\hat{B}^\dag\hat{B}},
\end{equation}
where $\Omega_{A(B)} (t)=\int_0^t dt'\,\omega_{A(B)} (t')$. The time evolution of the total density matrix is described by
\begin{equation}\label{R1}
\rho(t)=\hat{U}(t)\rho(0)\hat{U}^\dag,
\end{equation}
where $\rho(0)$ is typically chosen as a separable state $\rho_a (0)\otimes\rho_b(0)$. We are interested in the reduced density matrix $\rho_a (t)=tr_b(\rho(t))$ with components $_a\bra{n}\rho_a(t)\ket{m}_a$. In this scheme, the components of the reduced density matrix for the $a$-oscillator are obtained straightforwardly as (\ref{A})
\begin{eqnarray}\label{main}
&& _a\bra{n}\hat{\rho}_a(t)\ket{m}_a =\frac{1}{\sqrt{n!m!}}\sum_{s=0}^\infty\frac{(-1)^s}{s!}\,tr\Big([\hat{a}^\dag(t)]^{s+m}[\hat{a}(t)]^{s+n}\,\hat{\rho}(0)\Big),
\end{eqnarray}
where $\hat{a}(t)$ is in Heisenberg picture given by (\ref{A})
\begin{equation}\label{ahat}
\hat{a}(t)=e^{-i\omega_0 t}\cos[\kappa(t)]\,\hat{a}(0)-ie^{-i\omega_0 t}\sin[\kappa(t)]\,\hat{b}(0).
\end{equation}
For the particular case in which the $b$-oscillator is held in a thermal state, the initial state can be considered as the separable state
\begin{equation}\label{rozero}
\hat{\rho}(0)=\hat{\rho}_a (0)\otimes\frac{e^{-\beta_b\hbar\omega_0\hat{n}_b}}{z_b},
\end{equation}
where $z_b=tr(e^{-\beta_b\hbar\omega_0\hat{n}_b})$ is the partition function of the bath-oscillator. By inserting Eq.~(\ref{rozero}) into Eq.~(\ref{main}), one obtains (\ref{A})
\begin{eqnarray}\label{MainEq}
&& _a\bra{n}\hat{\rho}_a(t)\ket{m}_a =\nn\\
&& \frac{1}{\sqrt{n!m!}}\sum_{s=0}^\infty\frac{(-1)^s}{s!}\partial_{\lambda}^{s+m}\partial_{-\bar{\lambda}}^{s+n}
\Big[\chi_a (\lambda,\bar{\lambda},t)e^{-\lambda\bar{\lambda}\bar{n}_b\sin^2[\kappa(t)]}\Big]_{\lambda=\bar{\lambda}=0},
\end{eqnarray}
where
\begin{equation}
\chi_a (\lambda,\bar{\lambda},t)=tr_a\Big(e^{\lambda e^{i\omega_0 t}\cos[\kappa(t)]\hat{a}^\dag}e^{-\bar{\lambda}e^{-i\omega_0 t}\cos[\kappa(t)]\hat{a}}\hat{\rho}_a (0)\Big),
\end{equation}
is the normal characteristic function and $\bar{n}_{b}=1/(e^{\beta_{b}\hbar\omega_0}-1)$.
\subsection{The bath oscillator is held in zero temperature}
In this case, we set $\bar{n}_b=0$ in Eq.~(\ref{MainEq}), and obtain the time-evolution of the reduced density matrix for three different initial states: (i) thermal state (ii) coherent state, and (iii) number state, respectively as (\ref{B})
\begin{eqnarray}\label{RoTh}
&& \hat{\rho}_a (0)=\frac{e^{-\beta_a\hbar\omega_0\hat{n}_a}}{tr(e^{-\beta_a\hbar\omega_0\hat{n}_a})}\Rightarrow\nonumber\\
&& _a\bra{n}\hat{\rho}_a(t)\ket{m}_a=\delta_{nm}\,\frac{(\bar{n}_a\cos^2[\kappa(t)])^n}{(\bar{n}_a\cos^2[\kappa(t)]+1)^{n+1}},\nonumber\\
\end{eqnarray}
\begin{eqnarray}\label{RoCoh}
 \hat{\rho}_a (0)=\ket{\alpha}\bra{\alpha}&\Rightarrow&\nn\\
 _a\bra{n}\hat{\rho}_a(t)\ket{m}_a &=&\frac{(\alpha\cos[\kappa(t)])^n(\bar{\alpha}\cos[\kappa(t)])^m}{\sqrt{n!m!}},\nonumber\\
&=& _a\langle n|\underbrace{\alpha\cos[\kappa(t)]\rangle\langle\alpha\cos[\kappa(t)]}_{\hat{\rho}_a(t)}|m\rangle_a,
\end{eqnarray}
\begin{eqnarray}\label{RoNum}
&& \hat{\rho}_a (0)=\ket{N}\bra{N}\Rightarrow\nonumber\\
&& _a\bra{n}\hat{\rho}_a(t)\ket{m}_a=\delta_{nm}\,{{N}\choose {n}}(\cos^2[\kappa(t)])^n(\sin^2[\kappa(t)])^{N-n}.
\end{eqnarray}
\begin{figure}[H]
    \centering
    \includegraphics[width=.7\textwidth]{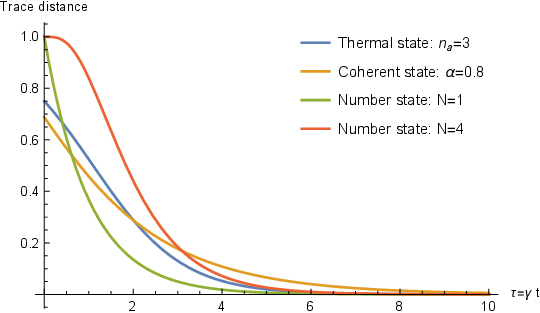}
    \caption{(Color online) Trace distance in terms of the dimensionless parameter $\tau=\gamma t$ for the choice $\cos^2[\kappa(t)]=e^{-\gamma t}$, and various initial states.}\label{fig1}
\end{figure}
\begin{figure}[H]
    \centering
    \includegraphics[width=.7\textwidth]{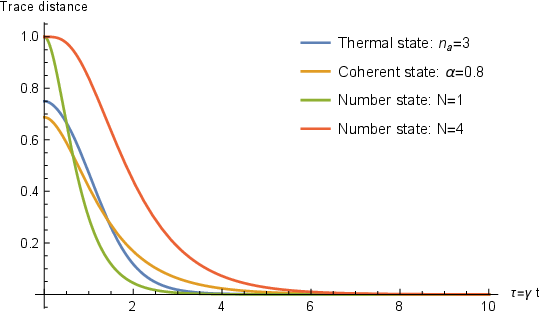}
    \caption{(Color online) Trace distance in terms of the dimensionless parameter $\tau=\gamma t$ for the choice $\cos^2[\kappa(t)]=\sin^2(\frac{\pi}{2}e^{-\gamma t})$, and various initial states.}\label{fig2}
\end{figure}
As a metric for measuring distances between states in Hilbert space, we use the trace distance $D_T (\cdot|\cdot)$ which is a monotonic measure defined by
\begin{equation}
D_T (\hat{\rho}_a(t)|\hat{\rho}_a(\infty))=D_T (t)=\frac{1}{2}tr(|\hat{\rho}_a (t)-\hat{\rho}_a (\infty)|),
\end{equation}
where $|\hat{O}|=\sqrt{\hat{O}^\dag \hat{O}}$. If operator $\hat{O}$ is Hermitian $\hat{O}^\dag=\hat{O}$, then $|\hat{O}|=\sum\limits_{i} |\lambda_i|$, which is a sum over the absolute value of the eigenvalues of $\hat{O}$. The trace distance is especially useful as a measure of the distinguishability of two quantum states. If $D_T (\rho|\sigma)=0$, then no measurement can distinguish $\rho$ from $\sigma$ \cite{Wilde2017}. It should be noted that the ME depends on the choice of the metric, but for monotonic measures, recent theoretical developments \cite{Tan2025}, show that all monotonic distance measures can be unified in describing ME.

The trace distance between the ground state $\hat{\rho}_a (\infty)=\ket{0}\bra{0}$ and the evolved state $\hat{\rho}_a (t)$ for three different types of initial states: (i) thermal, (ii) coherent, and (iii) number states are respectively given by (\ref{C})
\begin{eqnarray}\label{distances}
&& D^{\tiny{Th}}_T (t)=\frac{n_a\cos^2[\kappa(t)]}{n_a\cos^2[\kappa(t)]+1},\nonumber\\
&& D^{\tiny{Coh}}_T (t)=\sqrt{1-e^{-|\alpha|^2\cos^2[\kappa(t)]}},\nonumber\\
&& D^{\tiny{Num}}_T (t)=1-(\sin^2[\kappa(t)])^N.
\end{eqnarray}
In Figs.~(\ref{fig1},\ref{fig2}) the trace distances Eqs.~(\ref{distances}) are depicted in terms of the dimensionless variable $\tau=\gamma t$ for the choices $\cos^2[\kappa(t)]=e^{-\gamma t}$ and $\cos^2[\kappa(t)]=\sin^2 (\frac{\pi}{2}e^{-\gamma t})$, respectively. The behavior of curves in both figures are nearly identical and predict Mpemba effect.
\subsection{Coherent and Number states}
\begin{figure}[H]
    \centering
    \includegraphics[width=.7\textwidth]{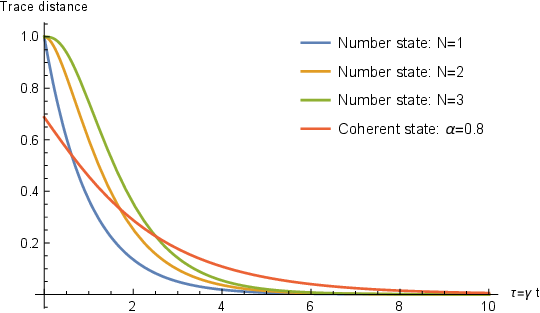}
    \caption{(Color online) The trace distance for coherent and number states in terms of $\tau=\gamma t$ for the choice $\cos^2[\kappa(t)]=e^{-\gamma t}$.}\label{fig3}
\end{figure}
In Fig.~\ref{fig3}, the temporal behavior of the trace distance is plotted for both coherent and number states. It can be observed that the number states intersect the coherent states, and this intersection occurs earlier for smaller values of
$N$. The initial energy of the oscillator in the Number states is proportional to $N$ and is greater than the initial energy in the coherent state, which is proportional to $|\alpha|^2$. Additionally, the initial distance of the numerical states from the equilibrium state $\hat{\rho}_a (\infty)=\ket{0}\bra{0}$ is larger than the initial distance of the coherent state from the equilibrium state. Therefore, the ME has occurred.
\subsection{Coherent and Thermal states}
\begin{figure}[H]
    \centering
    \includegraphics[width=.7\textwidth]{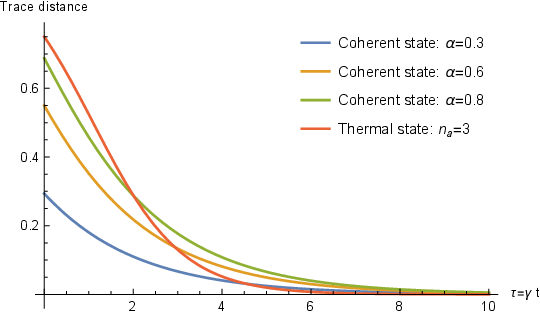}
    \caption{Color online) The trace distance for thermal and coherent states in terms of $\tau=\gamma t$ for the choice $\cos^2[\kappa(t)]=e^{-\gamma t}$.}\label{fig4}
\end{figure}
\begin{figure}[H]
    \centering
    \includegraphics[width=.7\textwidth]{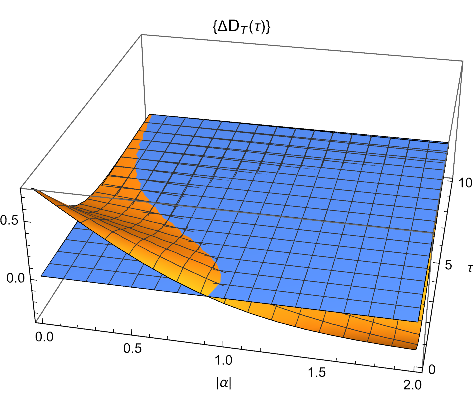}
    \caption{(Color online) The thermal state intersects the coherent states for the values of $|\alpha|$ where the difference $\Delta D_T (t)=D_T^{\tiny{Th}} (t)-D_T^{\tiny{Coh}} (t)$ changes sign. Here for $n_a=3$, we find approximately $0.3<|\alpha|<0.9$.}\label{fig5}
\end{figure}
In Fig.~\ref{fig4}, the temporal behavior of the trace distance is plotted for both thermal and coherent states. It can be observed that the thermal state intersects the coherent states. This intersection occurs for restricted values of $min\leq|\alpha|^2\leq max$, see Fig.~(\ref{fig5}). The initial energy of the oscillator in the coherent states is smaller than the initial energy in the thermal state. Additionally, the initial distance of the coherent states from the equilibrium state $\hat{\rho}_a (\infty)=\ket{0}\bra{0}$ is smaller than the initial distance of the thermal state from the equilibrium state. Therefore, the ME has occurred.
\subsection{Number and Thermal states}
\begin{figure}[h]
    \centering
    \includegraphics[width=.7\textwidth]{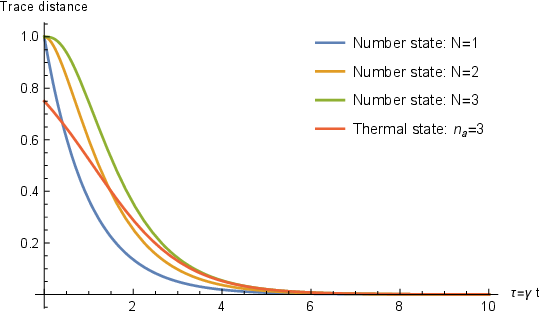}
    \caption{(Color online) The trace distance for thermal and number states in terms of $\tau=\gamma t$ for the choice $\cos^2[\kappa(t)]=e^{-\gamma t}$.}\label{fig6}
\end{figure}

In Fig.~\ref{fig6}, the temporal behavior of the trace distance is plotted for both thermal and number states. It can be observed that the number states intersect the thermal state for values $N< n_a$, and this intersection occurs earlier for smaller values of $N$. Though the initial distance of the thermal state from the equilibrium state $\hat{\rho}_a (\infty)=\ket{0}\bra{0}$ is smaller than the initial distance of the number states, the initial energy of the oscillator in the thermal states is smaller than the initial energy in the number states. Therefore, we may say that ME has occurred weakly.
\section{Dissipative two-level system}
\begin{figure}[H]
    \centering
    \includegraphics[width=.7\textwidth]{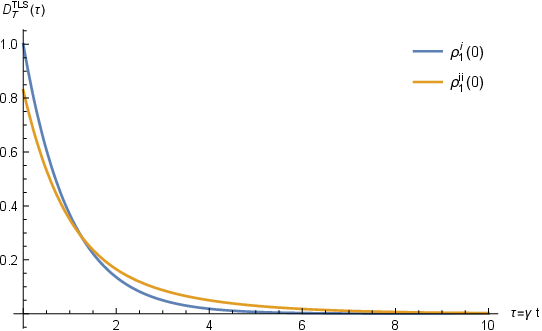}
    \caption{(Color online) The trace distance $D^{\tiny{TLS}}_T (t)$ in terms of the dimensionless parameter $\tau=\gamma t$ for the choice $\cos^2[\mu(t)]=e^{-\gamma t}$.}\label{fig7}
\end{figure}

\begin{figure}[H]
    \centering
    \includegraphics[width=.7\textwidth]{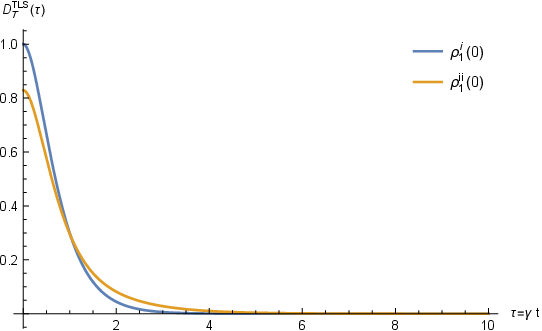}
    \caption{(Color online) The trace distance $D^{\tiny{TLS}}_T (t)$ in terms of the dimensionless parameter $\tau=\gamma t$ for the choice $\cos^2[\mu(t)]=\sin^2 (\frac{\pi}{2}e^{-\gamma t})$.}\label{fig8}
\end{figure}

A two-level system (TLS) is defined by two orthogonal states, a ground state $\ket{g}$ and an excited state $\ket{e}$, with corresponding energies $-\hbar\omega/2$ and $\hbar\omega/2$, respectively. Therefore, the Hamiltonian of a TLS can be written as
\begin{equation}
\hat{H}^{\tiny{TLS}}(t)=\frac{\hbar\omega}{2}\sigma_z,
\end{equation}
where
\begin{equation}
\sigma_z=\left(
           \begin{array}{cc}
             1 & 0\\
             0 & -1 \\
           \end{array}
         \right),
\end{equation}
is a Pauli matrix. The transition matrices are also defined by
\begin{equation}
\sigma_{+}=\ket{e}\bra{g}=\left(
           \begin{array}{cc}
             0 & 1\\
             0 & 0 \\
           \end{array}
         \right),\,\,\,\sigma_{-}=\ket{g}\bra{e}=\left(
           \begin{array}{cc}
             0 & 0\\
             1 & 0 \\
           \end{array}
         \right).
\end{equation}
To investigate a dissipative TLS, we need an auxiliary system which can be a copy of the main system. The auxiliary system can also be called a bath system. Both systems interact through a time-dependent coupling function $c(t)$. The total Hamiltonian can be written as
\begin{equation}
\hat{H}^{\tiny{TLS}}(t)=\frac{\hbar\omega}{2}\,(\sigma^1_z+\sigma^2_z)+\hbar\,c(t)\,(\sigma^1_{-}\sigma^2_{+}+\sigma^1_{+}\sigma^2_{-}).
\end{equation}
Let the main system be initially prepared in an arbitrary state $\hat{\rho}_1 (0)$ defined by parameters $r_x$, $r_y$, and $r_z$ with the constraint $r^2_x+r^2_y+r^2_z\leq 1$, and the bath system be initially prepared in a thermal state $\hat{\rho}_2 (0)$, then
\begin{eqnarray}
\rho(0)&=& \hat{\rho}_1 (0)\otimes\hat{\rho}_2 (0),\nonumber\\
&=& \left(
                                                  \begin{array}{cc}
                                                    (1+r_z)/2 & r_{-}/2 \\
                                                    r_{+}/2 & (1-r_z)/2 \\
                                                  \end{array}
                                                \right)\otimes\left(
                                                  \begin{array}{cc}
                                                    P_e & 0 \\
                                                    0 & P_g \\
                                                  \end{array}
                                                \right),\nonumber\\
       &=& \left(
                          \begin{array}{cccc}
                            \frac{(1+r_z)}{2}\,p_e & 0 & \frac{r_{-}}{2}\,p_e & 0 \\
                            0 & \frac{(1+r_z)}{2}\,p_g & 0 & \frac{r_{-}}{2}\,p_g \\
                            \frac{r_{+}}{2}\,p_e & 0 & \frac{(1-r_z)}{2}\,p_e & 0 \\
                            0 & \frac{r_{+}}{2}\,P_g & 0 & \frac{(1-r_z)}{2}\,p_g \\
                          \end{array}
                        \right),
\end{eqnarray}
where $r_{\mp}=r_x\mp i r_y$, and $P_e$ and $P_g$ are the thermal bath parameters with inverse temperature $\beta$
\begin{equation}
P_e=\frac{1}{1+e^{\beta\hbar\omega}},\,\,P_g=\frac{e^{\beta\hbar\omega}}{1+e^{\beta\hbar\omega}}.
\end{equation}
The components $\hat{\rho}_{1,ij} (t)$ of the reduced density matrix of the main system at an arbitrary time is given by \cite{Kh2024}
\begin{eqnarray}\label{evolvedtws}
 \hat{\rho}_{1,11} (t)=&& \frac{1+r_z}{2}P_e+\frac{1+r_z}{2}P_g\,\cos^2 [\mu(t)]\nonumber\\
                       &&+\frac{1-r_z}{2}P_e\sin^2 [\mu(t)],\nonumber\\
 \hat{\rho}_{1,12} (t)=&&\frac{r_{-}}{2}\,e^{-i\omega t}\cos[\mu(t)],\nonumber\\
 \hat{\rho}_{1,21} (t)=&&\frac{r_{+}}{2}\,e^{i\omega t}\cos[\mu(t)],\nonumber\\
 \hat{\rho}_{1,22} (t)=&&\frac{1-r_z}{2}P_g+\frac{1-r_z}{2}P_e\,\cos^2 [\mu(t)]\nonumber\\
                       &&+\frac{1+r_z}{2}P_g\sin^2 [\mu(t)],
\end{eqnarray}
where for convenience we defined $\mu(t)=\int_0^tdt'\,c(t')$. If the bath temperature is zero ($\beta=\infty$), then $P_e=0$ and $P_g=1$, therefore,
\begin{equation}\label{ground}
\hat{\rho}_2 (0)=\left(
                   \begin{array}{cc}
                     0 & 0 \\
                     0 & 1 \\
                   \end{array}
                 \right),
\end{equation}
and from Eqs.~(\ref{evolvedtws}) we obtain
\begin{equation}\label{es}
\hat{\rho}_1 (t)=\left(
                   \begin{array}{cc}
                     \frac{1+r_z}{2}\cos^2 [\mu(t)]  & \frac{r_{-}}{2}\,e^{-i\omega t}\cos[\mu(t)] \\
                     \frac{r_{+}}{2}\,e^{i\omega t}\cos[\mu(t)] & \frac{1-r_z}{2}+\frac{1+r_z}{2}\sin^2 [\mu(t)] \\
                   \end{array}
                 \right).
\end{equation}
The trace distance between the evolved state Eq.~(\ref{es}) and the relaxation state Eq.~(\ref{ground}) is (\ref{D})
\begin{equation}\label{DTLS}
D^{\tiny{TLS}}_T (t)=\frac{1}{2}\sqrt{(1+r_z)^2\cos^4[\mu(t)]+(r^2_x+r^2_y)\cos^2[\mu(t)]}.
\end{equation}
In Figs.~(\ref{fig7},\ref{fig8}), the trace distance between the evolved state Eq.~(\ref{es}) with the following initial states
\begin{eqnarray}
\label{initialTLS}
&& (r_x=0, r_y=0,r_z=1)\rightarrow\hat{\rho}^{i}_1 (0)=\left(
                                                  \begin{array}{cc}
                                                    1 & 0 \\
                                                    0 & 0 \\
                                                  \end{array}
                                                \right),\nonumber\\
&& (r_x=r_y=r_z=\frac{1}{2})\rightarrow\hat{\rho}^{ii}_1 (0)=\left(
                                               \begin{array}{cc}
                                                 3/4 & (1-i)/4 \\
                                                 (1+i)/4 & 1/4 \\
                                               \end{array}
                                             \right),
\end{eqnarray}
and the ground state Eq.~(\ref{ground}) have been depicted for both choices $\cos^2[\mu(t)]=e^{-\gamma t}$ and $\cos^2[\mu(t)]=\sin^2 (\frac{\pi}{2}e^{-\gamma t})$, respectively, indicating the Mpemba effect.
\subsection{Dissipative two-level system: another approach}
We can also couple the two-level system to a bosonic mode (oscillator) through a time-dependent coupling function $f(t)$. In this case, the Hamiltonian is
\begin{equation}\label{tlb1}
\hat{H} (t)=\frac{1}{2}\hbar\omega\,\sigma_z+\hbar\omega\,\hat{b}^\dag\hat{b}+\hbar f(t)\,(\hat{b}\sigma_{+}+\hat{b}^\dag\sigma_{-}),
\end{equation}
which is equivalent to the time-dependent Jaynes-Cummings model (TJCM) on resonance \cite{Gruver1993,Joshi1993,Schlicher1989,Gruver1994,Dasgupta1999}. This correspondence may provide a realisation of Mpemba effect in quantum optics.
In interaction picture, the Hamiltonian becomes
\begin{equation}\label{tlb2}
\hat{H}_I (t)=\hbar f(t)\,(\hat{b}\sigma_{+}+\hat{b}^\dag\sigma_{-}),
\end{equation}
which commutes at different times $[\hat{H}_I (t), \hat{H}_I (t')]=0$. Therefore, the evolution operator in interaction picture can be obtained as (\ref{E})
\begin{eqnarray}\label{tlb3}
\hat{U}_I (t) &&=e^{-i \phi(t) (\hat{b}\sigma_{+}+\hat{b}^\dag \sigma_{-})},\nonumber\\
              &&=\left(
                  \begin{array}{cc}
                    \cos\big(\phi(t)\sqrt{\hat{b}\hat{b}^\dag}\big) & -i\hat{b}\,\frac{\sin\big(\phi(t)\sqrt{\hat{b}^\dag\hat{b}}\big)}{\sqrt{\hat{b}^\dag\hat{b}}} \\
                    -i\hat{b}^\dag\frac{\sin\big(\phi(t)\sqrt{\hat{b}\hat{b}^\dag}\big)}{\sqrt{\hat{b}\hat{b}^\dag}} & \cos\big(\phi(t)\sqrt{\hat{b}^\dag\hat{b}}\big) \\
                  \end{array}
                \right),
\end{eqnarray}
where $\phi(t)=\int_0^t dt'\,f(t')$. The density matrix in interaction picture evolves as $\rho_I(t)=\hat{U}_I (t)\rho(0)\hat{U}^\dag_I (t)$. The density matrix in Schr\"{o}dinger picture is $\rho(t)=\hat{U}_0 (t)\rho_I (t)\hat{U}^\dag_0 (t)$, where
\begin{eqnarray}\label{uzero}
\hat{U}_0 (t)&=& e^{-i\frac{\omega t}{2}\sigma_z}\otimes e^{-i\omega t\hat{b}^\dag\hat{b}},\nonumber\\
             &=& \left(\begin{array}{cc}
         e^{\frac{-i\omega t}{2}}e^{-i\omega t\,\hat{b}^\dag\hat{b}} & 0 \\
            0 & e^{\frac{i\omega t}{2}}e^{-i\omega t\,\hat{b}^\dag\hat{b}} \\
               \end{array}
                \right).
\end{eqnarray}
The reduced density matrix of the main system (TLS) is obtained by performing a partial trace over the degrees of freedom of the bath-oscillator $\rho_s(t)=tr_b(\rho(t))$.
Let the system and the bath-oscillator be initially prepared in arbitrary states, then
\begin{eqnarray}\label{rozero22}
\rho(0) &=& \left(\begin{array}{cc}
         \frac{(1+r_z)}{2} & \frac{r_{-}}{2} \\
            \frac{r_{+}}{2} & \frac{(1-r_z)}{2} \\
               \end{array}
                \right)\otimes\hat{\rho}_b (0),\nonumber\\
        &=&  \left(\begin{array}{cc}
         \frac{(1+r_z)}{2}\,\hat{\rho}_b (0) & \frac{r_{-}}{2}\,\hat{\rho}_b (0) \\
            \frac{r_{+}}{2}\,\hat{\rho}_b (0) & \frac{(1-r_z)}{2}\,\hat{\rho}_b (0) \\
               \end{array}
                \right),
\end{eqnarray}
then by plugging Eqs.~(\ref{tlb3},\ref{uzero},\ref{rozero22}) into the equation
\begin{equation}\label{rhoreduced}
\rho_s(t)=tr_b \Big(\hat{U}_0 (t)\hat{U}_I (t)\rho(0)\hat{U}_I^\dag (t)\hat{U}_0^\dag (t)\Big),
\end{equation}
we obtain
\begin{eqnarray}
\rho_s(t)&=&tr_b \Big(\hat{U}_0 (t)\hat{U}_I (t)\rho(0)\hat{U}_I^\dag (t)\hat{U}_0^\dag (t)\Big),\nn\\
         &=&\left(
               \begin{array}{cc}
                 tr_b(\rho_{I,11}) & e^{-i\omega t}\,tr_b(\rho_{I,12}) \\
                 e^{i\omega t}\,tr_b(\rho_{I,21}) & tr_b(\rho_{I,22}) \\
               \end{array}
             \right),
\end{eqnarray}
where $\rho_{I,ij}=[\hat{U}_I (t)\rho(0)\hat{U}^\dag_I (t)]_{ij}$, and $\hat{U}_I (t)$ is defined by Eq.~(\ref{tlb3}). The explicit independent components of the reduced density matrix are
\begin{eqnarray}
\label{tdjc1}
\rho_{s,11}(t)&&= \frac{1+r_z}{2}\,tr_b\Big[\cos^2(\phi(t)\sqrt{\hat{n}+1})\hat{\rho}_b(0)\Big]\nn\\
&&\, +\frac{ir_{-}}{2}\,tr_b \Big[\frac{\sin(\phi(t)\sqrt{\hat{n}})}{\sqrt{\hat{n}}}\,\hat{b}^\dag\,\cos(\phi(t)\sqrt{\hat{n}+1})\hat{\rho}_b(0)\Big]\nn\\
&&\, -\frac{ir_{+}}{2}\,tr_b\Big[\cos(\phi(t)\sqrt{\hat{n}+1})\,\hat{b}\,\frac{\sin(\phi(t)\sqrt{\hat{n}})}{\sqrt{\hat{n}}}\hat{\rho}_b(0)\Big]\nn\\
&&\, +\frac{1-r_z}{2}\,tr_b\Big[\frac{\sin(\phi(t)\sqrt{\hat{n}})}{\sqrt{\hat{n}}}\,\hat{b}^\dag\hat{b}\,\frac{\sin(\phi(t)\sqrt{\hat{n}})}{\sqrt{\hat{n}}}\hat{\rho}_b(0)\Big],
\end{eqnarray}
\begin{eqnarray}
\label{tdjc2}
\rho_{s,12}(t)&&= e^{-i\omega t}\,\Big\{\frac{i(1+r_z)}{2}\,tr_b\Big[\frac{\sin(\phi(t)\sqrt{\hat{n}+1})}{\sqrt{\hat{n}+1}}\hat{b}\cos(\phi(t)\sqrt{\hat{n}+1})\hat{\rho}_b(0)\Big]\nn\\
&&\, +\frac{r_{-}}{2}\,tr_b \Big[\cos(\phi(t)\sqrt{\hat{n}})\cos(\phi(t)\sqrt{\hat{n}+1})\hat{\rho}_b(0)\Big]\nn\\
&&\, +\frac{r_{+}}{2}\,tr_b\Big[\frac{\sin(\phi(t)\sqrt{\hat{n}+1})}{\sqrt{\hat{n}+1}}\,\hat{b}^2\,\frac{\sin(\phi(t)\sqrt{\hat{n}})}{\sqrt{\hat{n}}}\hat{\rho}_b(0)\Big]\nn\\
&&\, -\frac{i(1-r_z)}{2}\,tr_b\Big[\cos(\phi(t)\sqrt{\hat{n}})\,\hat{b}\,\frac{\sin(\phi(t)\sqrt{\hat{n}})}{\sqrt{\hat{n}}}\hat{\rho}_b(0)\Big]\Big\}.
\end{eqnarray}
\subsubsection{TJCM and ME}
\begin{figure}[H]
    \centering
    \includegraphics[width=.7\textwidth]{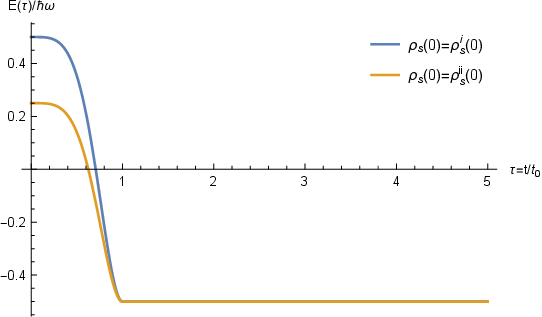}
    \caption{(Color online) The scaled energy of the two-level atom in units of $\hbar\omega$ for the initial states Eqs.~(\ref{initialTLS}) as a function of the dimensionless parameter $\tau=t/t_0$. }\label{fig9}
\end{figure}

\begin{figure}[H]
    \centering
    \includegraphics[width=.7\textwidth]{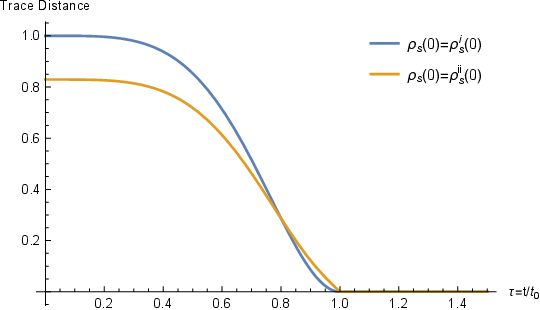}
    \caption{(Color online) The trace distances Eqs.~(\ref{DiDii1},\ref{DiDii2}) corresponding to the coupling function Eq.~(\ref{ramp}) in terms of the dimensionless time $\tau=t/t_0$. The ME has occurred after a finite scaled time around $\tau\approxeq 0.8$.}\label{fig10}
\end{figure}

\begin{figure}[H]
    \centering
    \includegraphics[width=.7\textwidth]{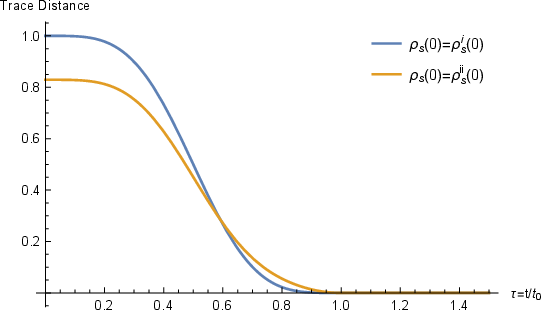}
    \caption{(Color online) The trace distances Eqs.~(\ref{DiDii1},\ref{DiDii2}) corresponding to the coupling function Eq.~(\ref{linearramp}) in terms of the dimensionless time $\tau=t/t_0$. The ME has occurred after a finite time around $t=0.6\,t_0$. }\label{fig11}
\end{figure}

\begin{figure}[H]
    \centering
    \includegraphics[width=.7\textwidth]{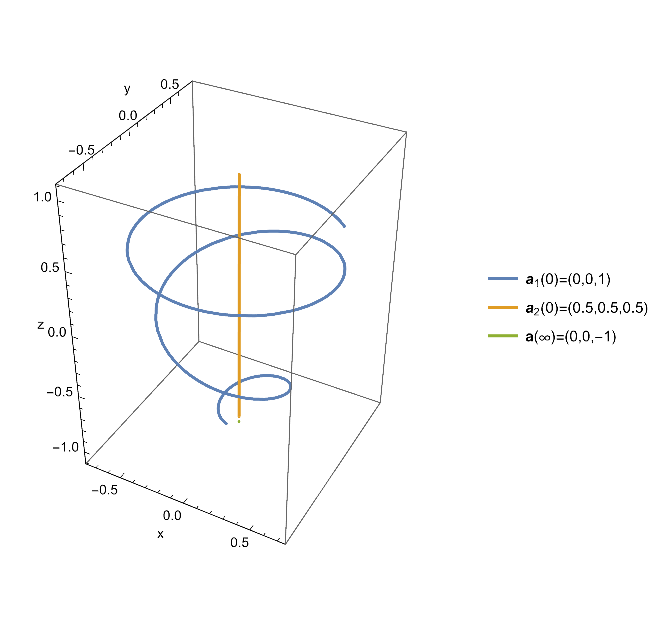}
    \caption{(Color online) The parametric plot of Eqs.~(\ref{veca}) for different initial states given by Eqs.~(\ref{initialTLS}), corresponding to $\mathbf{a}_1 (0)=(0,0,1)$ and $\mathbf{a}_2(0)=(0.5,0.5,0.5)$. The initially farther state $\mathbf{a}_1 (0)=(0,0,1)$ reaches the destination (relaxation point), corresponding to $\mathbf{a}(\infty)=(0,0,-1)$, sooner.} \label{fig12}
\end{figure}

\begin{figure}
    \centering
    \subfigure[$\tau\approx 0.57$]{\includegraphics[width=0.4\textwidth]{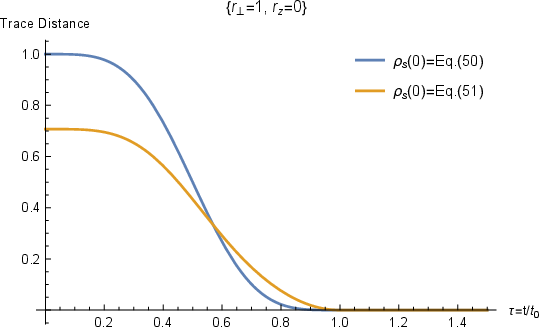}}
    \subfigure[$\tau\approx 0.63$]{\includegraphics[width=0.4\textwidth]{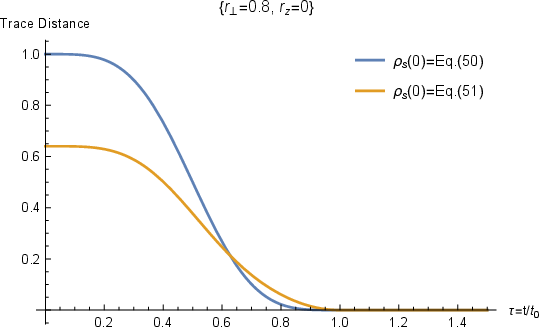}}
    \subfigure[$\tau\approx 0.68$]{\includegraphics[width=0.4\textwidth]{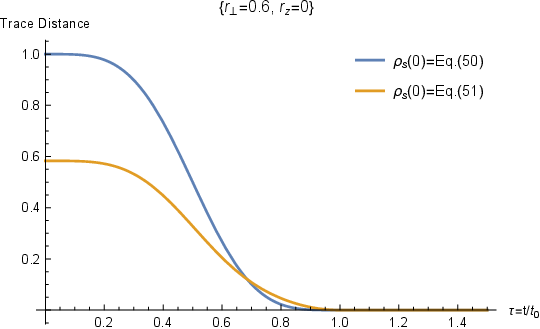}}
    \caption{The ME occurs at scaled times $\tau\approx 0.57,\,0.63,\,0.68$ which increases by decreasing the coherency.}
    \label{figs13}
\end{figure}

If the reservoir or the bath-oscillator is initially prepared in a thermal state $\hat{\rho}_b (0)=e^{-\beta\hbar\omega\hat{b}^\dag\hat{b}}/z_b$, with the partition function $z_b=tr(e^{-\beta\hbar\omega\hat{b}^\dag\hat{b}})$, then from Eqs.~(\ref{tdjc1},\ref{tdjc2}), we obtain
\begin{eqnarray}
\rho_{s,11}(t)&&=\frac{(1+r_z)}{2}\,\sum_{n=0}^\infty \cos^2(\phi(t)\sqrt{n+1})\,\frac{e^{-\beta\hbar\omega n}}{z_b}\nn\\
              &&\,+ \frac{(1-r_z)}{2}\,\sum_{n=0}^\infty \sin^2(\phi(t)\sqrt{n})\,\frac{e^{-\beta\hbar\omega n}}{z_b},\nn\\
\rho_{s,12}(t)&&=\frac{r_{-}}{2}\,e^{-i\omega t}\,\sum_{n=0}^\infty \cos(\phi(t)\sqrt{n+1})\cos(\phi(t)\sqrt{n})\,\frac{e^{-\beta\hbar\omega n}}{z_b},\nn\\
\rho_{s,21}(t)&&=\rho^*_{s,12} (t),\nn\\
\rho_{s,22}(t)&&=1-\rho_{s,11}(t).
\end{eqnarray}
In zero temperature, we have (appendix \ref{F})
\begin{equation}\label{roatzero}
\rho^{\tiny{(T=0)}}_s(t)=\left(\begin{array}{cc}
         \frac{(1+r_z)}{2}\cos^2[\phi(t)] & \frac{r_x-i r_y}{2}\,e^{-i\omega t}\cos[\phi(t)]\\
             \frac{r_x+ir_y}{2}\,e^{i\omega t}\cos[\phi(t)] & \frac{1-r_z}{2}+\frac{(1+r_z)}{2}\sin^2[\phi(t)]\\
               \end{array}
                \right).
\end{equation}
The density matrix Eq.~(\ref{roatzero}) can also be represented in terms of the time-dependent Bloch vector $\mathbf{a}(t)=(a_x (t),a_y (t),a_z (t))$ as
\begin{equation}\label{Bloch}
  \rho^{\tiny{(T=0)}}_s(t)=\frac{1}{2}(1+\mathbf{a}(t)\cdot\boldsymbol{\sigma}),
\end{equation}
where $\boldsymbol{\sigma}=(\sigma_x,\sigma_y,\sigma_z)$ are Pauli matrices. We have
\begin{eqnarray}
\label{veca}
  a_x (t) &=& r_x \cos[\omega t]\cos[\phi(t)]-r_y \sin[\omega t]\cos[\phi(t)],\nn \\
  a_y (t) &=& r_x \sin[\omega t]\cos[\phi(t)]+r_y \cos[\omega t]\cos[\phi(t)],\nn \\
  a_z (t) &=& r_z \cos^2[\phi(t)]-\sin^2[\phi(t)].
\end{eqnarray}
One can show that the trace distance between two density matrices $\hat{\rho}_1$ and $\hat{\rho}_2$ represented by respective Bloch vectors $\mathbf{a}_1$ and $\mathbf{a}_2$ is
\begin{equation}\label{Euclidean}
  D_{td}(\hat{\rho}_1|\hat{\rho}_2)=\frac{1}{2}\parallel\mathbf{a}_1-\mathbf{a}_2\parallel_2,
\end{equation}
where $\parallel\cdot\parallel_2$ is the Euclidean distance in $\mathbb{R}^3$. Now, considering the initial states given by Eqs.~(\ref{initialTLS}) for the two-level atom, the corresponding evolved density matrices are
\begin{eqnarray}
\label{evolvedstates}
  \rho_s^{i}(t)&=& \left(
                       \begin{array}{cc}
                         \cos^2[\phi(t)] & 0 \\
                         0 & \sin^2[\phi(t)] \\
                       \end{array}
                     \right),\,\,\,(r_x=r_y=0,\,r_z=1),\nn\\
 \rho_s^{ii}(t) &=& \left(
                       \begin{array}{cc}
                         \frac{3}{4}\cos^2[\phi(t)] & \frac{1-i}{4}\,e^{-i \omega t}\cos[\phi(t)] \\
                         \frac{1+i}{4}\,e^{i \omega t}\cos[\phi(t)] & \frac{1}{4}+\frac{3}{4}\sin^2[\phi(t)] \\
                       \end{array}
                     \right),\,(r_x=r_y=r_z=1/2).\nn\\
\end{eqnarray}
If the time-dependent coupling function is a ramp defined by \cite{Joshi1993}
\begin{equation}\label{ramp}
  f(t)=\left\{
         \begin{array}{ll}
           \frac{\pi t}{t_0^2}, & 0\leq t\leq t_0, \\
           0, & t\geq t_0,
         \end{array}
       \right.
\end{equation}
then
\begin{equation}\label{ramp2}
  \phi(t)=\left\{
            \begin{array}{ll}
              \frac{\pi}{2}\,(\frac{t}{t_0})^2, & 0\leq t \leq t_0, \\
              \frac{\pi}{2}, & t>t_0,
            \end{array}
          \right.
\end{equation}
and thermalization occurs in $t\rightarrow\infty$ independently of the initial state parameters $(r_z, r_{\pm})$
\begin{equation}\label{equilibrium}
\rho^{\tiny{(T=0)}}_s(\infty)=\left(\begin{array}{cc}
         0 & 0\\
             0 & 1\\
               \end{array}
                \right),
\end{equation}
as expected. The scaled energy of the two-level atom in units of $\hbar\omega$ has been depicted in Fig.~(\ref{fig9}) for the initial states Eqs.~(\ref{initialTLS}) as a function of the dimensionless parameter $\tau=t/t_0$.
The trace distance between the evolved state Eq.~(\ref{roatzero}) and the equilibrium state Eq.~(\ref{equilibrium}) is
\begin{equation}\label{tdgen}
  D_T (t)=\sqrt{\left(\frac{1+r_z}{2}\right)^2\cos^4[\phi(t)]+\frac{r^2_x+r^2_y}{4}\cos^2[\phi(t)]},
\end{equation}
therefore, the trace distance between the evolved states Eqs.~(\ref{evolvedstates}) and the equilibrium state Eq.~(\ref{equilibrium}) are
\begin{equation}\label{DiDii1}
  D_{T}^i(t) = \cos^2[\phi(t)],
\end{equation}
\begin{equation}\label{DiDii2}
  D_{T}^{ii}(t) = \frac{1}{4}\,\sqrt{9\cos^4[\phi(t)]+2\cos^2[\phi(t)]}.
\end{equation}
These trace distances are depicted in Fig.~(\ref{fig10}), indicating that ME occurs after a finite scaled time around $\tau\approxeq 0.8$ for the choice Eq.~(\ref{ramp}).
If a two-level atom with constant velocity $v$ passes trough a cavity with length $L$, and the shape function of the cavity mode is proportional to $\sin(\pi z/L)$, then the interaction term can be considered semiclassically as \cite{Schlicher1989}
\begin{equation}\label{linearramp}
  f_c(t)=\frac{\pi^2}{4 t_0}\,\sin(\pi t/t_0)\,\theta(t_0-t),
\end{equation}
where $t_0=L/v$ is the duration of interaction \cite{Haroche1989,Kimble1998,Koz2001}, $\theta(t)$ is the Heaviside step function, and we used $z=vt$. Here, we are not interested in the center-of-mass motion of the atom and will consider only the internal degrees of freedom. Now, we have
\begin{equation}\label{linearramp2}
  \phi(t)=\left\{
              \begin{array}{ll}
                \frac{\pi}{4}\,(1-\cos[\frac{\pi t}{t_0}]), & t\leq t_0 \\
                \frac{\pi}{2}, & t> t_0,
              \end{array}
            \right.
\end{equation}
In Fig.~(\ref{fig11}), the trace distances corresponding to the coupling function Eq.~(\ref{linearramp}) are plotted, showing that the ME occurs after a finite scaled time $\tau\approxeq 0.6$. For typical atom transit times $t_0\approx 10-100\,\mu s$, we have $t\approx 6-60\,\mu s$.
In Fig.~(\ref{fig12}), the parametric plots of Eqs.~(\ref{veca}) for different initial states given by Eqs.~(\ref{initialTLS}), corresponding to $\mathbf{a}_1 (0)=(0,0,1)$ and $\mathbf{a}_2 (0)=(0.5,0.5,0.5)$ a depicted in $\mathbb{R}^3$ in terms of the dimensionless time $t/t_0$. The initially farther state $\mathbf{a}_1 (0)=(0,0,1)$ reaches the destination (relaxation point), corresponding to $\mathbf{a} (\infty)=(0,0,-1)$, sooner.
In the general case, one can easily show that for the initial states
\begin{equation}\label{genInit1}
  \rho^{i}_s(o) = \left(
                       \begin{array}{cc}
                         1 & 0 \\
                         0 & 0 \\
                       \end{array}
                     \right),
\end{equation}
\begin{equation}\label{genInit2}
  \rho^{ii}_s(o) = \left(
                         \begin{array}{cc}
                           \frac{1+r_z}{2} & \frac{r_{-}}{2} \\
                           \frac{r_{+}}{2} &  \frac{1-r_z}{2} \\
                         \end{array}
                       \right),
\end{equation}
the corresponding trace distances from the equilibrium state in Eqs.~(\ref{DiDii1},\,\ref{tdgen}), intersect when $r_{\perp}=\sqrt{r^2_x+r^2_y}\neq 0$, and
\begin{equation}\label{findfi}
  \cos[\phi(t)]=\frac{r_{\perp}}{\sqrt{4-(1+r_z)^2}}.
\end{equation}
Therefore, a necessary condition for the occurrence of the ME in this case is the existence of initial coherence in Eq.~(\ref{genInit2}), which is proportional to $r_{\perp}^2$. For the maximal initial coherence,, $r_{\perp}=1$, we have $r_z=0$ and $\cos[\phi(t)]=1/\sqrt{3}$, leading to the intersection scaled time
\begin{equation}\label{tau}
  \tau=\frac{1}{\pi}\,\arccos\Big[1-\frac{4}{\pi}\arccos\big[\frac{r_{\perp}}{\sqrt{3}}\big]\Big].
\end{equation}
The trace distances for $r_z=0$ and different coherencies are depicted in Figs.~(\ref{figs13}), indicating that the intersection occurs at scaled times that increase as coherency decreases.
\subsubsection{Hilbert-Schmidt distance}
The Hilbert-Schmidt (HS) distance between two density operators $\rho$ and $\sigma$ is defined by
\begin{equation}\label{HS1}
  D_{HS} (\rho,\sigma)=\sqrt{tr[(\rho-\sigma)^2]}.
\end{equation}
The HS distance is easier to calculate but generally, it should not be employed as a distinguishability measure of quantum states \cite{Wilde2017}.

\begin{figure}[H]
    \centering
    \includegraphics[width=.6\textwidth]{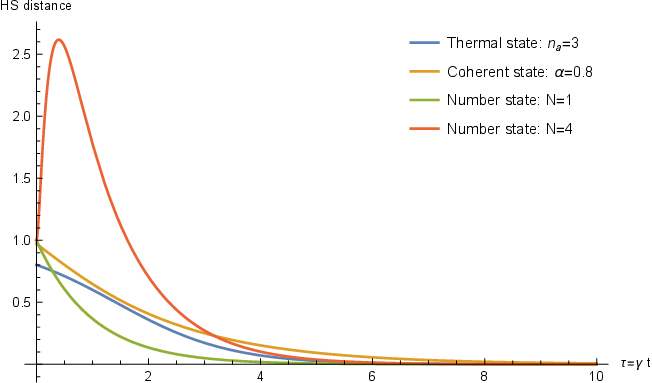}
    \caption{(Color online) The HS distance in terms of the dimensionless parameter $\tau=\gamma t$ for the choice $\cos^2[\kappa(t)]=e^{-\gamma t}$ and various initial states.} \label{fig14}
\end{figure}

\begin{figure}[H]
    \centering
    \includegraphics[width=.6\textwidth]{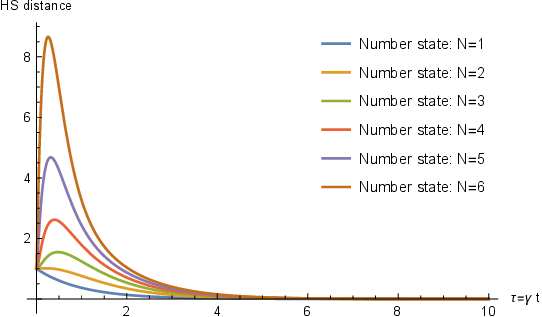}
    \caption{(Color online) The HS distance in terms of the dimensionless parameter $\tau=\gamma t$ for the choice $\cos^2[\kappa(t)]=e^{-\gamma t}$ and various number states. The HS distance is not a monotonic function for $n\geq 3$.} \label{fig15}
\end{figure}

If the oscillator is initially prepared in a coherent state $\hat{\rho}_a (0)=|\alpha\ra\la \alpha|$, then one can easily find (Eqs.~(\ref{distances}))
\begin{eqnarray}
   D_T(\hat{\rho}_a (t)|\hat{\rho}_a (\infty))=D^{Coh}_T (t)=\sqrt{1-e^{-|\alpha|^2\cos^2[\kappa(t)]}},\nn\\
  D_{HS}(\hat{\rho}_a (t)|\hat{\rho}_a (\infty))=D^{Coh}_{HS} (t)=\sqrt{2-2\,e^{-|\alpha|^2\cos^2[\kappa(t)]}}=\sqrt{2}\,D^{Coh}_T (t),
\end{eqnarray}
also, for a two-level system, the HS distance between the evolved state Eq.~(\ref{roatzero})
and the equilibrium state Eq.~(\ref{equilibrium}) is
\begin{eqnarray}\label{HSDiDii}
  D_{HS}(t) &=& \sqrt{2}\,\sqrt{\left(\frac{1+r_z}{2}\right)^2\cos^4[\phi(t)]+\frac{r^2_x+r^2_y}{4}\cos^2[\phi(t)]},\nn\\
            &=& \sqrt{2}\,D_{T}(t),
\end{eqnarray}
where we made use of Eq.~(\ref{tdgen}). Therefore, for an oscillator initially prepared in a coherent state and for a two-level system, the HS distances are proportional to the corresponding trace distances with the same proportionality factor $\sqrt{2}$; however, this does not apply in general. For example, consider a harmonic oscillator initially prepared in a number state $\hat{\rho}_a (0)=|N\ra\la N|$, interacting with its reservoir, we obtained the trace distance between the evolved state $\hat{\rho}_a (t)$ and the equilibrium state $\hat{\rho}_a (\infty)=|0\ra\la 0|$ as (Eqs.~(\ref{distances}))
\begin{equation}\label{DandH1}
  D_T(\hat{\rho}_a (t)|\hat{\rho}_a (\infty))=D^{Num}_T (t)=1-(\sin^2[\kappa (t)])^N,
\end{equation}
the Hilbert-Schmidt distance between the same density matrices can be obtained straightforwardly as
\begin{eqnarray}\label{DandH2}
&&  D_{HS}(\hat{\rho}_a (t)|\hat{\rho}_a (\infty))=D^{Num}_{HS} (t)=\sqrt{tr\big(\hat{\rho}_a (t)-\hat{\rho}_a (\infty)\big)^2}=\nn\\
&& \Big\{\sum_{n=1}^N \Big[{{N}\choose {n}} (\cos^2[\kappa(t)])^n (\sin^2[\kappa(t)])^{N-n}\Big]^2+\Big(1-(\sin^2[k(t)])^N\Big)^2\Big\}^{\frac{1}{2}}.\nn\\
\end{eqnarray}
As another example, let the oscillator be initially prepared in a thermal state $\hat{\rho}_a (0)=\frac{e^{-\beta_a\hbar\omega_0\hat{n}_a}}{z_a}$, then (Eqs.~(\ref{distances}))
\begin{eqnarray}\label{thermalHSD}
 && D^{\tiny{Th}}_T (t)=\frac{n_a\cos^2[\kappa(t)]}{n_a\cos^2[\kappa(t)]+1},\nonumber\\
 &&  D^{\tiny{Th}}_{HS} (t)=\sqrt{\frac{2n_a\cos^2[\kappa(t)]+2}{2n_a\cos^2[\kappa(t)]+1}}\,D^{\tiny{Th}}_T (t),
\end{eqnarray}
therefore, the HS distances are not necessarily a simple rescaling of the trace distances. Note that, in the long-time regime $\gamma t\gg 1$, the same factor $\sqrt{2}$, appears asymptotically in Eq.~(\ref{thermalHSD})
\begin{equation}\label{asym}
 \sqrt{\frac{2n_a\cos^2[\kappa(t)]+2}{2n_a\cos^2[\kappa(t)]+1}}=\sqrt{\frac{2n_a\,e^{-\gamma t}+2}{2n_a\,e^{-\gamma t}+1}} \rightarrow\,\sqrt{2}.
\end{equation}
In Fig.~(\ref{fig14}), the HS distances are depicted for various initial states. Again, the number state $|1\ra\la 1|$ evolves to the equilibrium state $|0\ra\la 0|$ faster. The HS distances for different number states are shown in Fig.~(\ref{fig15}), indicating that the HS distance for the number states with $n\geq 3$ is not a monotonic function of time.
\newpage

\appendix

\section{Derivation of Eqs.~(\ref{main},\ref{ahat},\ref{MainEq})}\label{A}
Using $\hat{\rho}_a (t)=tr_b(\hat{\rho} (t))$, we have
\beq\label{A1}
_a\bra{n}\hat{\rho}_a (t)\ket{m}_a=&& _a\bra{0}\frac{(\hat{a})^n}{\sqrt{n!}}\,\hat{\rho}_a (t)\,\frac{(\hat{a}^\dag)^m}{\sqrt{m!}}\ket{0}_a,\nn\\
                              =&& _a\bra{0}\,tr_b\big(\frac{1}{\sqrt{n! m!}}\,(\hat{a})^n \hat{\rho} (t) (\hat{a}^\dag)^m\big)\ket{0}_a,\nn\\
                              =&& tr_a \Big(\ket{0}_{aa}\bra{0}\,tr_b \Big[\frac{1}{\sqrt{n! m!}}\,(\hat{a})^n \hat{\rho} (t) (\hat{a}^\dag)^m\Big]\Big).
\eeq
by inserting the identity
\be\label{A2}
\ket{0}_{aa}\bra{0}=\sum_{s=0}^\infty \frac{(-1)^s}{s!}\,(\hat{a}^\dag)^s (\hat{a})^s,
\ee
into Eq.~(\ref{A1}), we obtain
\beq\label{A3}
_a\bra{n}\hat{\rho}_a (t)\ket{m}_a=&& tr\Big[\frac{1}{\sqrt{n! m!}}\sum_{s=0}^\infty \frac{(-1)^s}{s!}\,(\hat{a}^\dag)^s (\hat{a})^s\,(\hat{a})^n \hat{\rho} (t) (\hat{a}^\dag)^m\Big],\nn\\
                                  =&& \frac{1}{\sqrt{n! m!}}\sum_{s=0}^\infty \frac{(-1)^s}{s!}\,tr\Big((\hat{a}^\dag)^{s+m} (\hat{a})^{s+n}\,\hat{\rho} (t)\Big).
\eeq
Now, using $\hat{\rho} (t)=\hat{U} (t)\hat{\rho} (0) \hat{U}^\dag (t)$, and the Heisenberg representations
\beq\label{A4}
\hat{a} (t)=\hat{U}^\dag (t) \hat{a} \hat{U} (t),\nn\\
\hat{a}^\dag (t)=\hat{U}^\dag (t) \hat{a}^\dag \hat{U} (t),
\eeq
we rewrite Eq.~(\ref{A3}) as
\be\label{A5}
_a\bra{n}\hat{\rho}_a (t)\ket{m}_a=\frac{1}{\sqrt{n! m!}}\sum_{s=0}^\infty \frac{(-1)^s}{s!}\,tr\Big([\hat{a}^\dag (t)]^{s+m} [\hat{a} (t)]^{s+n}\,\hat{\rho} (0)\Big).
\ee

By making use of the Bogoliubov transformations Eq.~(\ref{Bogo}), the Hamiltonian Eq.~(\ref{U1}), and definitions $\Omega_{A(B)} (t)=\omega_0 t\mp \kappa(t)$, ($\kappa(t)=\int_0^t dt'\,g(t')$), we obtain
\beq\label{A6}
\hat{a} (t)=&& \underbrace{\frac{1}{2}\big(e^{-i\Omega_B (t)}+e^{-i\Omega_A (t)}\big)}_{c_1 (t)=e^{-i\omega_0 t}\cos(\kappa(t))} \hat{a}+ \underbrace{\frac{1}{2}\big(e^{-i\Omega_B (t)}-e^{-i\Omega_A (t)}\big)}_{c_2 (t)=-ie^{-i\omega_0 t}\sin(\kappa(t))}\hat{b}.
\eeq
We can rewrite Eq.~(\ref{A5}) as
\beq\label{A7}
&& _a\bra{n}\hat{\rho}_a (t)\ket{m}_a=\frac{1}{\sqrt{n! m!}}\sum_{s=0}^\infty \frac{(-1)^s}{s!}\,tr\Big(\partial^{s+m}_\lambda e^{\lambda\hat{a}^\dag (t)}\partial^{s+n}_{-\bar{\lambda}} e^{-\bar{\lambda}\hat{a}(t)}\,\hat{\rho} (0)\Big),\nn\\
&&                                  = \frac{1}{\sqrt{n! m!}}\sum_{s=0}^\infty \frac{(-1)^s}{s!}\partial^{s+m}_{\lambda}\partial^{s+n}_{-\bar{\lambda}}\big(\chi_a (\lambda,\bar{\lambda},t)\,\chi_b (\lambda,\bar{\lambda},t)\big),
\eeq
where we defined
\beq
&&\chi_a (\lambda,\bar{\lambda},t)=tr_a\Big(e^{\lambda e^{i\omega_0 t}\cos[\kappa(t)]\hat{a}^\dag}
                                  e^{-\bar{\lambda} e^{-i\omega_0 t}\cos[\kappa(t)]\hat{a}}\hat{\rho}_a (0)\Big),\nn\\
&&\chi_b (\lambda,\bar{\lambda},t)=tr_b\Big(e^{i\lambda e^{i\omega_0 t}\sin[\kappa(t)]\hat{b}^\dag}
                                  e^{i\bar{\lambda} e^{-i\omega_0 t}\sin[\kappa(t)]\hat{b}}\,\frac{e^{-\beta_b\hbar\omega_0 \hat{n}_b}}{z_b}\Big),
\eeq
To obtain Eq.~(\ref{A7}), we inserted Eq.~(\ref{A6}) and its conjugate into Eq.~(\ref{A5}), and made use of the initial state
\beq\label{sini}
\hat{\rho} (0)=\hat{\rho}_a (0)\otimes\underbrace{\frac{e^{-\beta_2\hbar\omega_0\,\hat{n}_b}}{z_b}}_{\hat{\rho}_b (0)}.
\eeq
Now by expanding the exponential terms in the last term of Eq.~(\ref{A7}) and using the identity
\beq\label{A9}
tr_b \Big[(\hat{b}^\dag)^p \,\hat{b}^q\,\hat{\rho}_b (0)\Big]&=& \frac{1}{z_b}\sum_{k=0}^{\infty}\, _b\bra{k} (\hat{b}^{\dag})^p (\hat{b})^q\ket{k}_b \,e^{-\beta_2\hbar\omega k},\nn\\
                                                             &=&  \frac{\delta_{pq}}{z_b}\,\sum_{k=0}^{\infty}\, _b\bra{k} (\hat{b}^\dag)^q (\hat{b})^q\ket{k}_b \,e^{-\beta_2\hbar\omega k},\nn\\
                                                             &=&  \frac{\delta_{pq}}{z_b}\,\sum_{k=q}^\infty \frac{k!}{(k-q)!}\,e^{-\beta_2\hbar\omega k},\nn\\
                                                             &=&  \frac{\delta_{pq}}{z_b}\,\sum_{k=0}^\infty \frac{(k+q)!}{k!}\,e^{-\beta_2\hbar\omega (k+q)},\nn\\
                                                             &=&  \delta_{pq}\,q!\,\bar{n}^q_b,\,\,\,\,\,(\bar{n}_b=\frac{1}{e^{\beta_2\hbar\omega}-1}),
\eeq
we obtain
\begin{eqnarray}\label{A10}
&& _a\bra{n}\hat{\rho}_a(t)\ket{m}_a =\nn\\
&& \frac{1}{\sqrt{n!m!}}\sum_{s=0}^\infty\frac{(-1)^s}{s!}\partial_{\lambda}^{s+m}\partial_{-\bar{\lambda}}^{s+n}
\Big[tr_a\Big(\chi_a (\lambda,\bar{\lambda},t)\Big)e^{-\lambda\bar{\lambda}\bar{n}_b\sin^2[\kappa(t)]}\Big]_{\lambda=\bar{\lambda}=0},\nn\\
\end{eqnarray}
\section{Derivation of Eqs.~(\ref{RoTh},\ref{RoCoh},\ref{RoNum})}\label{B}
\begin{eqnarray}\label{B1}
_a\bra{n}\hat{\rho}_a(t)\ket{m}_a =\frac{1}{\sqrt{n!m!}}\sum_{s=0}^\infty\frac{(-1)^s}{s!}\partial_{\lambda}^{s+m}\partial_{-\bar{\lambda}}^{s+n}
\Big[tr_a\Big(\chi_a (\lambda,\bar{\lambda},t)\Big)\Big]_{\lambda=\bar{\lambda}=0}.\nn\\
\end{eqnarray}
For a thermal state $\hat{\rho}_a (0)=e^{-\beta_a\hbar\omega_0\hat{n}_a}/z_a$, where $z_a=tr_a (e^{-\beta_a\hbar\omega_0\hat{n}_a})$ is the partition function. In this case, following the same steps we did in deriving Eq.~(\ref{A9}), we find
\beq\label{B2}
tr_a(\chi_a (\lambda,\bar{\lambda},t)) &=& tr_a\Big(e^{\lambda e^{i\omega_0 t}\cos[\kappa(t)]\hat{a}^\dag}e^{-\bar{\lambda}e^{-i\omega_0 t}\cos[\kappa(t)]\hat{a}}\hat{\rho}_a (0)\Big),\nonumber\\
                                        &=& e^{-\lambda\bar{\lambda}\cos^2[\kappa(t)]\bar{n}_a}.
\eeq
Therefore,
\begin{eqnarray}\label{B3}
_a\bra{n}\hat{\rho}_a(t)\ket{m}_a &=& \frac{1}{\sqrt{n!m!}}\sum_{s=0}^\infty\frac{(-1)^s}{s!}\partial_{\lambda}^{s+m}\partial_{-\bar{\lambda}}^{s+n}
\Big[e^{-\lambda\bar{\lambda}\cos^2[\kappa(t)]\bar{n}_a}\Big]_{\lambda=\bar{\lambda}=0},\nn\\
                                  &=& \frac{1}{\sqrt{n!m!}}\sum_{s=0}^\infty\frac{(-1)^s (s+n)!}{s!}\delta_{nm}\big(\bar{n}_a \cos^2[\kappa(t)]\big)^{s+n},\nn\\
                                  &=& \delta_{nm}\,\frac{\big(\bar{n}_a \cos^2[\kappa(t)]\big)^n}{\big(\bar{n}_a \cos^2[\kappa(t)]+1\big)^{n+1}},
\end{eqnarray}
where we used the identity
\be\label{B4}
\sum_{s=0}^\infty \frac{(-1)^s (s+n)!}{s!}(x)^{s}=\frac{n!}{(1+x)^{n+1}}.
\ee
For a coherent state $\hat{\rho}_a (0)=|\alpha\rangle\langle \alpha|$, we obtain
\begin{eqnarray}\label{B5}
&& tr_a\Big(e^{\lambda e^{i\omega_0 t}\cos[\kappa(t)]\hat{a}^\dag}e^{-\bar{\lambda}e^{-i\omega_0 t}\cos[\kappa(t)]\hat{a}}|\alpha\rangle\langle \alpha|\Big)\nn\\
&& = \langle \alpha|e^{\lambda e^{i\omega_0 t}\cos[\kappa(t)]\hat{a}^\dag}e^{-\bar{\lambda}e^{-i\omega_0 t}\cos[\kappa(t)]\hat{a}} |\alpha\rangle,\nn\\
&& = e^{\lambda\bar{\alpha}e^{i\omega_0 t}\cos[\kappa(t)]}e^{-\bar{\lambda}\alpha e^{-i\omega_0 t}\cos[\kappa(t)]}.
\end{eqnarray}
From Eq.~(\ref{B1}), we find
\begin{eqnarray}\label{B6}
&& _a\bra{n}\hat{\rho}_a(t)\ket{m}_a = \nn\\
&& \frac{1}{\sqrt{n!m!}}\sum_{s=0}^\infty\frac{(-1)^s}{s!}(\bar{\alpha} e^{i\omega_0 t}\cos[\kappa(t)])^{s+m}(\alpha e^{-i\omega_0 t}\cos[\kappa(t)])^{s+n},\nn\\
&&= \frac{1}{\sqrt{n!m!}}(\bar{\alpha} e^{i\omega_0 t}\cos[\kappa(t)])^{m}(\alpha e^{-i\omega_0 t}\cos[\kappa(t)])^{n}\,e^{-|\alpha|^2\cos^2[\kappa(t)]},\nn\\
&&= _a\langle n|\alpha e^{-i\omega_0 t}\cos[\kappa(t)]\rangle\langle \alpha e^{-i\omega_0 t}\cos[\kappa(t)]|m\rangle_a,
\end{eqnarray}
therefore, $\hat{\rho}_a (0)$ evolves to the following coherent state
\begin{equation}\label{B7}
\hat{\rho}_a (t)=|\alpha e^{-i\omega_0 t}\cos[\kappa(t)]\rangle\langle \alpha e^{-i\omega_0 t}\cos[\kappa(t)]|.
\end{equation}
For a number state $\hat{\rho}_a (0)=\ket{N}_{aa}\bra{N}$, we have
\begin{eqnarray}
&& tr_a\Big(e^{\lambda e^{i\omega_0 t}\cos[\kappa(t)]\hat{a}^\dag}e^{-\bar{\lambda}e^{-i\omega_0 t}\cos[\kappa(t)]\hat{a}}\ket{N}_{aa}\bra{N}\Big)=\nn\\
&& \bra{N}e^{\lambda e^{i\omega_0 t}\cos[\kappa(t)]\hat{a}^\dag}e^{-\bar{\lambda} e^{-i\omega_0 t}\cos[\kappa(t)]\hat{a}}\ket{N},\nn\\
&&= \sum_{r,s=0}^\infty \frac{(\lambda e^{i\omega_0 t}\cos[\kappa(t)])^r}{r!}\frac{(-\bar{\lambda}e^{-i\omega_0 t}\cos[\kappa(t)])^s}{s!}\underbrace{\bra{N}(\hat{a}^\dag)^r(\hat{a})^s\ket{N}}_{\delta_{rs}\frac{N!}{(N-r)!}},\nn\\
&&= \sum_{r=0}^N \frac{N!}{r! (N-r)!}\,(-|\lambda|^2\cos^2[\kappa(t)])^r=(1-\lambda\bar{\lambda}\cos^2[\kappa(t)])^N.
\end{eqnarray}
Therefore,
\begin{eqnarray}\label{B8}
_a\bra{n}\hat{\rho}_a(t)\ket{m}_a &=& \frac{1}{\sqrt{n!m!}}\sum_{s=0}^\infty\frac{(-1)^s}{s!}\partial^{s+m}_\lambda\partial^{s+n}_{-\bar{\lambda}}\Big(1-\lambda\bar{\lambda}\cos^[\kappa(t)]\Big)_{\lambda=\bar{\lambda}=0},\nn\\
                                  &=& \frac{\delta_{nm}}{n!}\sum_{s=0}^{N-n}\frac{(-1)^s}{s!}\frac{N!}{(N-s-n)!}(\cos^2[\kappa(t)])^{s+n},\nn\\
                                  &=& \delta_{nm} \frac{N!}{n! (N-n)!}(\cos^2[\kappa(t)])^n \underbrace{\sum_{s=0}^{N-n}{{N-n}\choose {s}} (\cos^2[\kappa(t)])^s}_{(1-\cos^2[\kappa(t)])^{N-n}},\nn\\
                                  &=& \delta_{nm} {{N}\choose {n}}\big(\cos^2[\kappa(t)]\big)^n \big(\sin^2[\kappa(t)]\big)^{N-n}.
\end{eqnarray}
\section{Derivation of Eqs.~(\ref{distances})}\label{C}
For a thermal state, using Eq.~(\ref{B3}) and the definition of trace distance, we have
\begin{eqnarray}
D^{\tiny{Th}}_T (t) &=& \frac{1}{2}\left(\sum_{n=1}^\infty \frac{(\bar{n}_a \cos^2[\kappa(t)])^n}{(\bar{n}_a \cos^2[\kappa(t)]+1)^{n+1}}+\Big(1-\frac{1}{\bar{n}_a \cos^2[\kappa(t)]+1}\Big)\right),\nn\\
                    &=& \frac{1}{2}\left(1-\frac{1}{\bar{n}_a \cos^2[\kappa(t)]+1}+1-\frac{1}{\bar{n}_a \cos^2[\kappa(t)]+1}\right),\nn\\
                    &=& \frac{\bar{n}_a \cos^2[\kappa(t)]}{\bar{n}_a \cos^2[\kappa(t)]+1}.
\end{eqnarray}
For a coherent state, using Eq.~(\ref{B7}) and the identity \cite{Wilde2017}
\begin{equation}
\frac{1}{2}tr(|\ket{\psi}\bra{\psi}-\ket{\phi}\bra{\phi}|)=\sqrt{1-|\la\psi|\phi\ra|^2},
\end{equation}
we obtain
\begin{eqnarray}
D^{\tiny{Coh}}_T (t) &=& \sqrt{1-|\langle{\alpha e^{-i\omega_0 t}\cos[\kappa(t)]}|0\rangle|^2},\nn\\
                     &=& \sqrt{1-e^{-|\alpha|^2\cos^2[\kappa(t)]}}.
\end{eqnarray}
For a number state, using Eq.~(\ref{B8}), we have
\begin{eqnarray}
D^{\tiny{Num}}_T (t) &=& \frac{1}{2}\sum_{n=1}^N {{N}\choose {n}}\big(\cos^2[\kappa(t)]\big)^n \big(\sin^2[\kappa(t)]\big)^{N-n}\nn\\
                     && +\frac{1}{2}\Big(1-(\sin^2[\kappa(t)])^N\Big),\nn\\
                     &=& \frac{1}{2}\Big(1-(\sin^2[\kappa(t)])^N+1-(\sin^2[\kappa(t)])^N\Big),\nn\\
                     &=& 1-(\sin^2[\kappa(t)])^N.
\end{eqnarray}
\section{Derivation of equations Eq.~(\ref{DTLS})}\label{D}
From Eqs.~(\ref{ground},\ref{es}), we have
\begin{equation}
\hat{\rho}_1 (t)-\hat{\rho}_2 (0)=\left(
                   \begin{array}{cc}
                     \frac{1+r_z}{2}\cos^2 [\mu(t)]  & \frac{r_{-}}{2}\,e^{-i\omega t}\cos[\mu(t)] \\
                     \frac{r_{+}}{2}\,e^{i\omega t}\cos[\mu(t)] & -\frac{1+r_z}{2}\cos^2 [\mu(t)] \\
                   \end{array}
                 \right),
\end{equation}
with eigenvalues
\begin{equation}
\lambda_{\pm}=\pm\frac{1}{2}\sqrt{(1+r_z)^2\,\cos^4[\mu(t)]+(r_x^2+r_y^2)\,\cos^2[\mu(t)]},
\end{equation}
therefore,
\begin{eqnarray}
D^{\tiny{TLS}}_T (t) &=& \frac{1}{2}(|\lambda_+|+|\lambda_-|),\nn\\
                     &=& \frac{1}{2}\sqrt{(1+r_z)^2\cos^4[\mu(t)]+(r^2_x+r^2_y)\cos^2[\mu(t)]}.
\end{eqnarray}
\section{Derivation of Eq.~(\ref{tlb3})}\label{E}
We have
\begin{eqnarray}
i\hbar \frac{d}{dt}\hat{U}_I (t)=-i\hbar f(t)\left(
                   \begin{array}{cc}
                    \sqrt{\hat{b}\hat{b}^\dag}\sin [\phi(t)\sqrt{\hat{b}\hat{b}^\dag}]  & i\hat{b}\cos[\phi(t)\sqrt{\hat{b}^\dag\hat{b}}] \\
                     i\hat{b}^\dag\cos[\phi(t)\sqrt{\hat{b}\hat{b}^\dag}] & \sqrt{\hat{b}^\dag\hat{b}}\sin [\phi(t)\sqrt{\hat{b}^\dag\hat{b}}] \\
                   \end{array}
                 \right),\nn\\
\end{eqnarray}
on the other hand
\begin{eqnarray}
&& \hat{H}_I (t)\hat{U}_I (t)=\nn\\
&& \left(
                   \begin{array}{cc}
                    0  & \hbar f(t)\,\hat{b} \\
                    \hbar f(t)\,\hat{b}^\dag & 0 \\
                   \end{array}
                 \right)\left(
                  \begin{array}{cc}
                    \cos\big(\phi(t)\sqrt{\hat{b}\hat{b}^\dag}\big) & -i\hat{b}\,\frac{\sin\big(\phi(t)\sqrt{\hat{b}^\dag\hat{b}}\big)}{\sqrt{\hat{b}^\dag\hat{b}}} \\
                    -i\hat{b}^\dag\frac{\sin\big(\phi(t)\sqrt{\hat{b}\hat{b}^\dag}\big)}{\sqrt{\hat{b}\hat{b}^\dag}} & \cos\big(\phi(t)\sqrt{\hat{b}^\dag\hat{b}}\big) \\
                  \end{array}
                \right),\nonumber\\
                        =&& -i\hbar f(t) \left(
                   \begin{array}{cc}
                    \sqrt{\hat{b}\hat{b}^\dag}\sin [\phi(t)\sqrt{\hat{b}\hat{b}^\dag}]  & i\hat{b}\cos[\phi(t)\sqrt{\hat{b}^\dag\hat{b}}] \\
                     i\hat{b}^\dag\cos[\phi(t)\sqrt{\hat{b}\hat{b}^\dag}] & \sqrt{\hat{b}^\dag\hat{b}}\sin [\phi(t)\sqrt{\hat{b}^\dag\hat{b}}] \\
                   \end{array}
                 \right),
\end{eqnarray}
therefore,
\begin{equation}
i\hbar \frac{d}{dt}\hat{U}_I (t)=\hat{H}_I (t)\hat{U}_I (t).
\end{equation}
\section{Derivation of Eq.~(\ref{roatzero})}\label{F}
In zero temperature we have $\hat{\rho}_b (0)=\ket{0}\bra{0}$, therefore,
\begin{eqnarray}
&& \langle\cos^2[\phi(t)\sqrt{\hat{b}\hat{b}^\dag}]\rangle_B=\langle 0|\cos^2[\phi(t)\sqrt{\hat{n}_b+1}]|0\rangle=\cos^2[\phi(t)],\nonumber\\
&& \langle\cos^2[\phi(t)\sqrt{\hat{b}^\dag\hat{b}}]\rangle_B=\langle 0|\cos^2[\phi(t)\sqrt{\hat{n}_b}]|0\rangle=1,\nonumber\\
&& \langle\sin^2[\phi(t)\sqrt{\hat{b}\hat{b}^\dag}]\rangle_B=\langle 0|\sin^2[\phi(t)\sqrt{\hat{n}_b+1}]|0\rangle=\sin^2[\phi(t)],\nonumber\\
&& \langle\sin^2[\phi(t)\sqrt{\hat{b}^\dag\hat{b}}]\rangle_B=\langle 0|\sin^2[\phi(t)\sqrt{\hat{n}_b}]|0\rangle=0,\nonumber\\
&& \langle\cos[\phi(t)\sqrt{\hat{b}^\dag\hat{b}}]\cos[\phi(t)\sqrt{\hat{b}\hat{b}^\dag}]\rangle_B =\langle 0|\cos[\phi(t)\sqrt{\hat{n}_b}]\cos[\phi(t)\sqrt{\hat{n}_b+1}]|0\rangle,\nn\\
&& =\cos[\phi(t)],
\end{eqnarray}
leading to Eq.~(\ref{roatzero}).
\newpage

\noindent\text{\textbf{Declaration of competing interest}}\\

The authors declare that they have no known competing financial interests or personal relationships that could have appeared to influence the work reported in this paper.\\

\noindent\text{\textbf{Acknowledgements}}\\

This work has been supported by the University of Kurdistan. The authors thank Vice Chancellorship
of Research and Technology, University of Kurdistan.\\

\noindent\text{\textbf{Data availability}}\\

No data was used for the research described in the article.


\end{document}